\documentclass[aps, prl, reprint,amsmath,amssymb, longbibliography]{revtex4-1}
\usepackage{amsmath }
\usepackage{graphicx}
\usepackage{dcolumn}
\usepackage{natbib}
\usepackage{float}
\begin{document}
\title{Quantum diffraction of position-momentum entangled photons from a sharp edge}
\author{Samridhi Gambhir and Mandip Singh}
\email{mandip@iisermohali.ac.in}
\affiliation{Department of Physical Sciences,\\
Indian Institute of Science Education and Research Mohali, Mohali, 140306, India.}
\begin{abstract}
  In this paper, an experiment of quantum diffraction of position-momentum entangled photons from a straight sharp edge is presented. Path of a single photon of an entangled pair is partially blocked by a sharp edge whereas the other photon is detected at a stationary location without revealing the which-path information of the other photon. Quantum diffraction pattern of the sharp edge is revealed only in the correlated conditional detection of spatially separated photons and no diffraction pattern is formed in local detections of individual photons. Theoretical analysis of the quantum diffraction of position-momentum entangled photons from a sharp edge is also presented in this paper. Experimental measurements of the quantum diffraction pattern are compared with theoretically calculated quantum diffraction pattern of position-momentum entangled photons. 

\end{abstract}
\maketitle

 Interference of a single particle with itself is a true manifestation of principle of quantum superposition. A single photon exhibits interference in a double slit experiment, if no which-path information of its passage through a double slit is available. If this photon is quantum entangled with another photon such that it carries the which-path information of the former photon, then a single photon interference is not formed. However, an interference pattern can be recovered selectively from a smeared pattern by making measurements  of photon quantum state in a path superposition basis and correlating it with detection of the other photon on the screen.  In this type of conditional and selective detection of individual photons, interference manifests a global nature of quantum entanglement by which amplitudes of both photons interfere, even if only one of them has passed through the double slit. This feature of quantum entanglement has been demonstrated in quantum ghost interference experiments \cite{ghost1, ghost2, ghost3, ghost4, zeirev1}. In quantum ghost interference selection is naturally linked to location of detection of a photon. Based on coincidence detection, experiments of ghost imaging have been performed \cite{gimage1, gimage2, gimage3, gimage4, gimage5, gimage6}. Quantum interference is also shown in an experiment consisting of a delocalized double slit \cite{ndslit, cond}. Similar experiments considering EPR \cite{epr} pairs of entangled atoms have been proposed \cite{zei}. From a foundational perspective of quantum physics, interesting thought provoking experiments have been reported on quantum erasure of path information showing global interference of quantum amplitudes of entangled photons \cite {scullyqer, qer1, qer2, qer3, qer4, zeirev2, scully2,popper}.  
 
In this paper, an experiment on quantum diffraction of position-momentum entangled photons from a straight sharp edge is presented. Diffraction of classical electromagnetic waves is known since the beginning of Huygens-Fresnel theory of diffraction however, a first rigorous theory of diffraction of classical light from an edge was given by A. Sommerfeld \cite{sommer}. Diffraction of classical light from a straight sharp edge \cite{bornwolf, edge1,edge2, optik,edge3}, often regarded as a precursor of diffraction, has been intensively explored.
 
In an experiment presented in this paper, a single photon from a pair of position-momentum entangled photons is interacted with a sharp edge and detected by a stationary single photon detector. While the other photon, from the same pair of entangled photons, is detected at a fixed location by another single photon detector.  Diffraction pattern of a sharp edge is obtained in correlated conditional measurements of photon counts by moving the sharp edge. This experiment shows a global quantum diffraction effect of position-momentum entangled photons in correlated conditional measurements. However, no diffraction pattern is formed in separate local detections of photons. Quantum diffraction pattern depends on the optical distance between detectors and a sharp edge and it is independent of location of an infinitely extended source. A displacement of detectors produces a shift in the quantum diffraction pattern. Furthermore, a theoretical analysis of quantum diffraction is given in this paper. Correlated conditional diffraction patterns measured in the experiment are in agreement with theoretical calculations of quantum diffraction patterns of a straight sharp edge. 

\section{Diffraction of position-momentum entangled photons from a sharp edge}
\begin{figure}
\begin{center}
\includegraphics[scale=0.105]{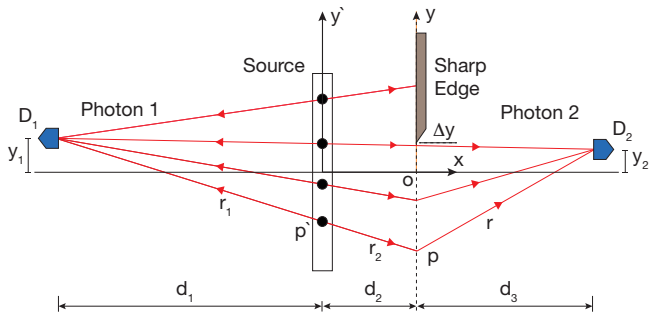}
\caption{\label{fig1} \emph{Position-momentum entangled photons are produced by an extended source with pair production amplitude $\psi(y')$. Photon 1 is detected by $D_{1}$, photon 2 is partially blocked by a sharp edge and detected by $D_{2}$.}}
\end{center}
\end{figure}

To findout quantum diffraction pattern of position-momentum entangled photons from a straight sharp edge, consider a schematic diagram given in Fig.~\ref{fig1}.  In experiment, position-momentum entangled photons are produced by a type-I spontaneous parametric down conversion (SPDC) process happening in a second-order nonlinear crystal, such as Beta Barium Borate (BBO). Similar SPDC sources of photon pairs have been utilised in the quantum ghost interference experiments \cite{ghost1}.  In experiments, photon pair creation rate is kept low so that a single photon pair is produced and detected. Probability of having more than one photon pairs in one experimental cycle is extremely low.  In this paper, the analysis of quantum diffraction is given in two-dimensions of space that is according to the experimental setup of the source as explained further. 

Type-I SPDC process produces photon pairs at a nonzero angle in general \emph{w.r.t} momentum direction of pump photons to conserve momentum and energy. Momenta of photons of each pair, produced in SPDC process, are opposite to each other in a plane perpendicular to the direction of propagation of pump photons. For such an extended source, photons of from a pair are position-momentum entangled in a transverse two-dimensional plane. Therefore, a two-dimensional configuration is considered further. 
For an extended SPDC source, a pair of photons is produced at a same location in crystal however, location of pair production is delocalized \emph{i.e.} amplitude of pair production of photons is delocalized over the extension of the source.  Consider an arbitrary point $p'$ in the source such that photons from a pair produced around a point $p'$ have opposite directions of their momentum. If an exact location of pair production is determined then an uncertainty in momentum of each photon becomes so large that it results in a separable quantum state of photons.  Consider, $|\bf{p}_{1}\rangle_{y'}$ is a momentum quantum state of a photon 1, produced around $p'(y')$, going towards a detector $D_{1}$ and $|\bf{p}_{2}\rangle_{y'}$ is the momentum quantum state of photon 2, produced around $p'(y')$ in the opposite direction to the momentum of photon 1. Each photon has same energy and momentum changes with direction of propagation. Their exact location of origin is uncertain in the vicinity of an arbitrary  point $p'(y')$. If an amplitude of a pair production at an arbitrary location $y'$ in the source is $\psi(y')$, then quantum state of both photons of same polarization can be written as
\begin{equation}
\label{eq:1}
|\Psi\rangle \propto \int \psi(y')|\bf{p}_{1}\rangle_{y'}|\bf{p}_{2}\rangle_{y'} \mathrm{d}y'
\end{equation}

Directions of momenta $\bf{p}_1$ and $\bf{p}_2$ of photons are opposite to each other. Quantum state given in Eqn.~\ref{eq:1} is a two-photon position-momentum entangled state under a consideration that a photon 1 is going towards detector 1. Both photons are momentum entangled as well as they are position entangled in their location of origin. Amplitude of pair production is normalized, such that $\int^{\infty}_{-\infty}\psi^{\ast}{(y')}\psi{(y')} \mathrm{d}y'=1$. 
Photon 1 can be detected by a single photon detector $D_{1}$ whereas the path of photon 2 is partially blocked by a straight sharp edge located on $y$-axis, which is parallel to $y'$-axis. One end of the sharp edge is positioned at $\Delta y$ and the other end is assumed to be positioned at infinity on $y$-axis. After traversing the sharp edge, photon 2 can be detected by another single photon detector $D_{2}$. Both the detector, $D_{1}$ and $D_{2}$, are positioned at fixed locations $y_{1}$ and $y_{2}$, respectively, parallel to $y$-axis. Distance along $x$-axis of detector $D_{1}$ from the sharp edge is $d_{1}+d_{2}$ and distance of detector $D_{2}$ from the sharp edge is $d_{3}$. Since extension of source is large therefore, from a detection of photon 1 at detector $D_{1}$ one cannot find out even in principle from which location in the source the photon pair is emitted. Therefore, a detection of a single photon is not revealing which-path information of the other photon. Directions of momenta of photons are opposite therefore, if a pair is originated around a point $p'(y')$ and photon 1 is detected at location $y_{1}$, then photon 2 is most likely to be at a point $p(y)$ on $y$-axis such that $y'=y_{1}+(y-y_{1})d_{1}/(d_{1}+d_{2})$. Detection of photons is not revealing any information about the momenta of photons emitted by an extended source. Once a photon 1 is detected at a known location and due to nonlocal quantum state reduction \cite{epr} a particular amplitude to find a second photon at $p(y)$ on $y$-axis containing a sharp edge can be immediately determined. This selection is made by a position resolved detection of photon 1.Therefore, a particular amplitude of photon 2 can be written as 
\begin{equation}
\label{eq:2}
a(y) \propto \frac{\psi(y')}{(2\pi \hbar)^{2}} e^{i p_{1} r_{1}/\hbar} e^{i p_{2} r_{2}/\hbar}
\end{equation}
 where $p_{1}$ and $p{_2}$ are magnitudes of momentum of photon 1 and photon 2 respectively. Distance between a detector $D_{1}$ and an arbitrary point $p'(y')$ is $r_{1}$. Distance of a point $p(y)$ from the arbitrary point $p'(y')$ is $r_{2}$ as shown in Fig.~\ref{fig1}.  The amplitude $a(y)$ is a complex function in general, it varies along $y$-axis containing a sharp edge.  Now, due to diffraction from the sharp edge, there are different possible paths via which a photon 2 can reach at detector $D_{2}$. To find out an amplitude of detection of photon 2 at detector $D_{2}$, if photon 1 is detected by a detector $D_{1}$, the amplitude of photon 2 emanating from a point $p$ on $y$-axis can be considered as a circular wave originating from a point $p$. If photon 1 is not detected by $D_{1}$ then this event is not counted as a conditional event and in this case the particular amplitude $a(y)$ is not well defined. Paths of photon 2 are partially blocked by a sharp edge on $y$-axis. The circular wave amplitudes from different locations on $y$-axis are quantum superimposed once a photon 1 is detected at location $y_{1}$ by a detector $D_{1}$ as the which-path information of photon 2 is not revealed by detection of photon 1. Total amplitude of a position correlated conditional detection, $a_{12}$, of both photons can be evaluated by superimposing two-photon amplitudes in the unblocked region such that
 \begin{equation}
\label{eq:3}
a_{12} \propto \frac{1}{(2\pi\hbar)^{2}}\int^{\Delta y}_{-\infty} \psi(y') e^{i k_{o}r_{1}} e^{i k_{o}r_{2}} \frac{e^{ik_{o}r}}{r^{1/2}} \mathrm{d}y
\end{equation}

where a magnitude of momentum of each photon is $\hbar k_{o}$, $k_{o}=2 \pi/\lambda$ is a wave vector magnitude corresponding to wavelength $\lambda$ and $r$ is a distance of a detector $D_{2}$ from a point $p$ on $y$-axis. Integrand function is a circular wave amplitude $e^{ik_{o}r}/r^{1/2}$ multiplied by $a(y)$.  In experimental situations the extension of the source is finite. The first term $\psi(y')$ of integral in Eqn.~\ref{eq:3} is an amplitude of a pair production that vanishes at infinity. Actual size of the source is determined by the transverse extension of pump laser beam that leads to a finite width of $\psi(y')$. Therefore, a finite size of the source is due to a finite extension of $\psi(y')$. 

From Fig.~\ref{fig1}, total distance is $r_{1}+r_{2}=((d_{1}+d_{2})^{2}+(y-y_{1})^{2})^{1/2}$. Since a location of detector 1 is close to $x$-axis and its distance from source is much larger as compared to the extension of the source therefore, total distance can be expressed as  $r_{1}+r_{2}\simeq(d_{1}+d_{2})(1+(y-y_{1})^{2}/2(d_{1}+d_{2})^2)$. Similarly, a distance of detector $D_{2}$ from an arbitrary point $p$ on $y$-axis can be written as $r\simeq d_{3}(1+(y-y_{2})^{2}/2d^{2}_{3})$. Here, a contribution of term $e^{i k_{o}r}/r^{1/2}$ to integral in Eqn.~\ref{eq:3} diminishes as point $p$ is moved away from origin $o$. In addition, an amplitude of a pair production in the crystal is determined by the transverse field profile of pump laser beam. For a Gaussian transverse mode of pump laser the amplitude of a pair production is real and it can be written as $ \psi(y')\propto e^{-y'^{2}/\sigma^{2}}$, where $\sigma$ is width of a pair production amplitude. The amplitude $\psi(y')$ of a pair production becomes smaller and smaller as magnitude of $y'$ increases and eventually it vanishes at infinity. Therefore, total amplitude of correlated conditional detection of photons can be evaluated as
\begin{equation}
\label{eq:4}
a_{12}=\frac{c_{n}e^{i k_{o} (d_{1}+d_{2}+d_{3})}}{(2\pi\hbar)^{2} d^{1/2}_{3}}\int^{\Delta y}_{-\infty} e^{-\frac{y'^{2}}{\sigma^{2}}} e^{i k_{o}\frac{(y-y_{1})^{2}}{2(d_{1}+d_{2})}} e^{ik_{o}\frac{(y-y_{2})^{2}}{2d_{3}}}  \mathrm{d}y
\end{equation}
where $c_{n}$ is a constant and $y'=y_{1}+(y-y_{1})d_{1}/(d_{1}+d_{2})$. Therefore, a probability of conditional detection of photons can be calculated as $p_{12}= a^{\ast}_{12} a_{12}$. Probability of conditional detection of photons for different position $\Delta y$, of a sharp edge produces a quantum diffraction pattern of the sharp edge. In experimental situation, probability of correlated conditional detection of photons is proportional to counts of correlated coincidence detection of photons in a chosen time interval. In experiment, position of each detector is stationary when a sharp edge is moved. If any detector is displaced then complete coincidence diffraction pattern is shifted. The correlated coincidence diffraction pattern given in Eqn.~\ref{eq:4} depends on distances of a sharp edge from individual detectors and it is independent of location of an infinitely extended source.  It is also evident that there is no diffraction pattern formation by individual photons \emph{i.e.} photon counts of each detector show no diffraction pattern as a sharp edge is moved. Quantum diffraction pattern is formed in correlated conditional measurements of photon counts only. Quantum diffraction pattern is formed even if a sharp edge is placed in the path of the other quantum entangled photon and it is independent of the order of detection of photons.

\section{Experimental realization}
\begin{figure}
\begin{center}
\includegraphics[scale=0.082]{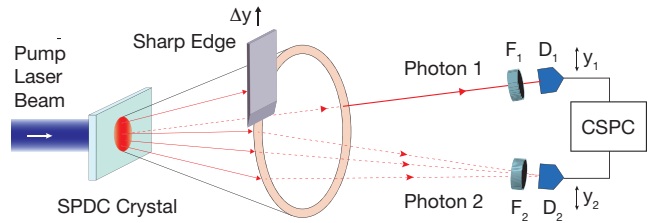}
\caption{\label{fig2} \emph{ A schematic diagram of experiment to measure quantum diffraction pattern of position-momentum entangled photons from a sharp edge.}}
\end{center}
\end{figure}

Experiment of quantum diffraction is performed with position-momentum entangled photons of same energy as shown in Fig.~\ref{fig2}. Quantum entangled photons are produced by SPDC happening in a second-order nonlinear BBO crystal. Phase matching angle of the crystal is chosen for type-I phase matching of degenerate SPDC.

\begin{figure*}[p]
\begin{center}
\includegraphics[scale=0.72]{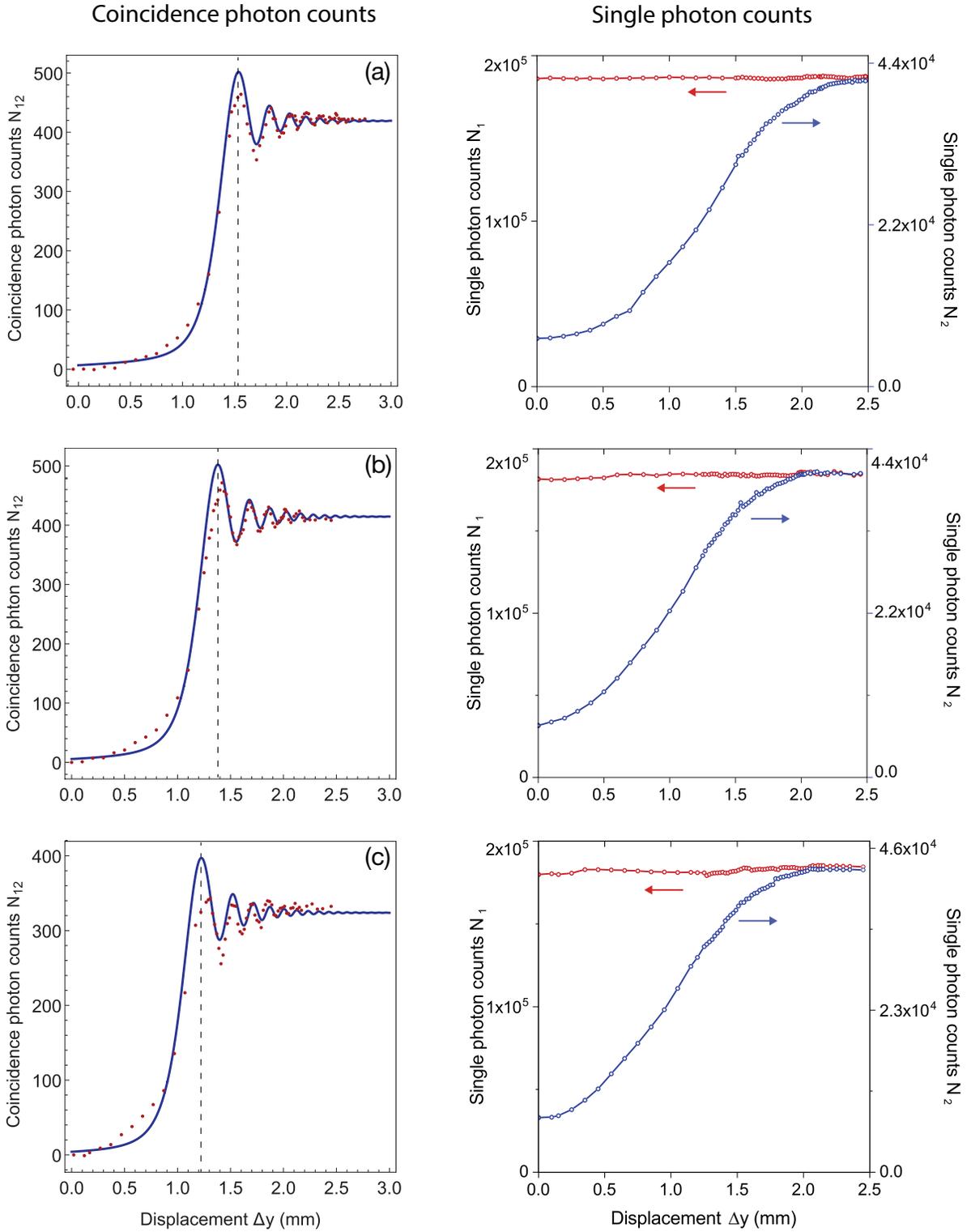}
\caption{\label{fig3} \emph{Quantum diffraction patterns of a sharp edge for different detector location is given in the left column. Diffraction pattern is shifted as detector $D_{2}$ is displaced. Detector location is (a) $y_{2}=1.52~mm$, (b) $y_{2}=1.32~mm$ and (c) $y_{2}= 1.12~mm$.Each single point represents experimentally measured position correlated coincidence photon counts for 30~sec integration time. Each measured  pattern is compared with the quantum diffraction pattern calculated from equation 4 and shown by a solid line plot. Plots given in the right column are the corresponding single photon counts of each detector.}}
\end{center}
\end{figure*}

A pump laser beam of wavelength 405~$nm$ is incident on the crystal and pairs of position-momentum entangled photons are produced. These photon pairs are quantum entangled in a two-dimensional transverse plane \emph{w.r.t} the direction of propagation of pump beam.
It is important to note that correlated coincidence amplitude depends on the exponents in the integral shown in Eqn.~\ref{eq:4}, that are in the form of a ratio of photon momentum to distances $d_{1}+ d_{2}$ and $d_{3}$. All such quantities are defined in two-dimensions. However, in a real experiment, these quantities can be measured in three dimensions and their respective projections in a transverse plane can be placed in the integral. The projection factors of momentum and distance in their ratio term cancel each other. Therefore,  photon momentum and distances indicated further in the paper are measured in three-dimensions. Wavelength of each down-converted photon is about 810~$nm$. Pump laser beam diameter is expanded to produce position-momentum entangled pairs of photons over a broader transverse extension of the crystal. Width $\sigma$ of pair production amplitude $\psi(y')$ increases with pump laser beam diameter. In this way, number of observable fringes of a quantum diffraction pattern can also be increased. Quantum diffraction pattern is independent of rate of pair production, if photon pairs are resolved in time. Each photon is detected by fiber coupled single photon detectors. Photons are coupled to optical fibers and fiber couplers are placed on three-dimensional translation stages. A straight sharp edge is placed to partially block photon 2 going towards detector $D_{2}$. Each single photon detector is preceded by optical bandpass filters ($F_{1}$ and $F_{2}$) of center wavelength 810~$nm$. Single photon counts and correlated conditional single photon counts of both detectors are measured with a two channel coincidence single photon counter (CSPC). A measurement time window of CSPC is chosen such that a single pair of entangled photons is resolved in time.

To measure quantum diffraction pattern of a sharp edge,  position $\Delta y$  of a sharp edge is displaced in steps. For each displacement, single photon counts and coincidence single photon counts are measured. Locations, $y_{1}$ and $y_{2}$ of single photon detectors $D_{1}$ and $D_{2}$ are stationary. Experiment is repeated for three different locations, $y_{2}$, of detector $D_{2}$. Quantum diffraction pattern of a sharp edge is shown in left column of Fig.~\ref{fig3} for three different locations of detector $D_{2}$. Corresponding single photon counts of each detector are shown in the right column of Fig.~\ref{fig3}. Location $y_{1}=0.15~mm$ of detector $D_{1}$, and distances $d_{1}=50~cm$ , $d_{2}=28~cm$ and $d_{3}=22~cm$ are measured in three-dimensions and they are same in all plots. Each experimental data point representing coincidence photon counts $N_{12}$ is acquired for 30~$sec$ integration time. Solid line in each coincidence count plot is corresponding to a quantum diffraction pattern calculated by solving Eqn.~\ref{eq:4}, where $\sigma=0.85~mm$. 
Quantum diffraction pattern of a sharp edge is revealed only in correlated coincidence photon counts. It is evident from experiment and theory results shown in  Fig.~\ref{fig3} that the quantum diffraction pattern is shifted with the displacement of position, $y_{2}$, of a single photon detector $D_{2}$. Quantum diffraction pattern also shows a shift with displacement of position, $y_{1}$, of a single photon detector $D_{1}$.

In single photon measurements by a detector, there is no way to know at what location the other photon is detected or has it been detected by a detector or not. This lack of information suppresses the formation of diffraction pattern in the local single photon counts. If a photon 1 is undetected then a particular amplitude, Eqn.~\ref{eq:2}, is undefined.  In other words, the phase distribution of photon 2 amplitude in a plane of a sharp edge is not well defined and it suppresses the formation of a diffraction pattern in single local measurements of photons. Once a photon 1 is detected at a known location the phase of amplitude of photon 2 in the plane of the sharp edge is immediately known due to nonlocal quantum state reduction that is a consequence of quantum entanglement. In this way, this quantum state reduction process leads to a formation of quantum diffraction pattern in correlated coincidence photon detection measurements. In other words, a position correlated coincidence detection selects a particular phase distribution of a two-photon amplitude that results in a formation of a quantum diffraction pattern. On the other hand, If only coincidences are measured without position correlation of photon 1 then no diffraction pattern is formed even in coincidence measurements because of a random phase fluctuation in shot to shot measurement. It has been shown that the quantum diffraction pattern is independent of location of an infinitely extended source. If experiment can store the information of detection of photons then detection events can be correlated pairwise and quantum diffraction pattern can also be recovered even after completing the experiment.

{\bf{Contribution of Authors}}: This experiment is conceived by Mandip Singh (MS). MS made the theoretical model of quantum diffraction from a sharp edge. Samridhi Gambhir (SG) has taken the data and plotted patterns.  MS made diagrams and explained the quantum diffraction pattern. Both authors discussed  the experiment. MS wrote the manuscript.

\bibliography{ref}


\end{document}